# Attitudes Toward Ambiguity Among Self-employed and Incorporated Entrepreneurs

Thomas Åstebro

Frank M. Fossen

Cédric Gutierrez[*]

Abstract

How do entrepreneurs act on their beliefs when probabilities of outcomes are unknown but subjectively perceived? We theorize that two distinct dimensions of ambiguity attitudes influence entrepreneurial action: ambiguity aversion—the unwillingness to bear ambiguity—and ambiguity sensitivity—how individuals discriminate between different levels of perceived chances of success. The second dimension determines how much entrepreneurs adjust their actions based on new information—a distinct aspect that cannot be captured by ambiguity aversion alone. Our theory suggests that entrepreneurs with different growth orientations have different ambiguity attitudes as compared to employees. Using incentivized measures from a large-scale survey, we find that incorporated entrepreneurs exhibit lower ambiguity aversion than employees, indicating that they are more willing to act under ambiguity. Distinctively, unincorporated self-employed individuals show higher ambiguity sensitivity, indicating that their actions are more responsive to changes in their beliefs. These patterns persist after controlling for risk attitudes, optimism, cognitive ability, and demographics. Our results highlight the distinct impacts of ambiguity aversion and ambiguity sensitivity on entrepreneurial actions.

---

[*] Thomas Åstebro, astebro@hec.fr, HEC Paris, 1 rue de la Liberation, 78350 Jouy-en-Josas, France. Frank M. Fossen (corresponding author), ffossen@unr.edu, University of Nevada-Reno, 1664 N. Virginia Street, Reno, NV 89557, USA. Cédric Gutierrez, cedric.gutierrez@unibocconi.it, Department of Management and Technology and ICRIOS, Bocconi University, Via Roentgen, 1, 20136, Milano, Italy. We thank Christian Pugnaghi Zimpelmann for valuable comments, as well as Raghuram Jonnalagedda, Ke Lyu and Trevor McLemore for excellent research assistance. Frank Fossen thanks HEC Paris, where he conducted a part of this research as a visiting scholar, and University of Nevada, Reno, for research sabbatical assistance.

# 1. Introduction

Frank Knight (1921) distinguished uncertainty, where probabilities of outcomes cannot be objectively determined, from routine business risk where probabilities are known, arguing that entrepreneurial profit arises from willingness to act under uncertainty. While there is still a debate about what exactly constitutes Knightian uncertainty, this insight captures a fundamental aspect of entrepreneurship: entrepreneurs form subjective beliefs about events and act on these beliefs, despite not knowing if they are right. What drives entrepreneurial action is therefore not just what entrepreneurs believe will happen but also their attitude toward ambiguity, which reflects how beliefs lead to action when probabilities are unknown.[1] This willingness to bear ambiguity represents a key element of entrepreneurial action theories (McMullen and Shepherd, 2006; Foss and Klein, 2012); yet, current approaches often treat ambiguity attitudes as unidimensional, overlooking key distinctions identified in decision science between different components of how individuals respond to ambiguity.

Consider two prospective entrepreneurs. The first is deciding whether to start a neighborhood food truck. Because this is an established business opportunity, the entrepreneur can rely more on observable data about foot traffic, local demographic data, competitors' menu pricing, and historical failure rates. The second entrepreneur is deciding whether to launch a deep-tech venture, developing a new sodium-ion battery. Here, the mapping from today's lab results and manufacturing principles to tomorrow's commercial performance is far less clear. Both entrepreneurs have to rely primarily on subjective probabilities and, therefore, face ambiguity, but the level of ambiguity is higher for the deep-tech entrepreneur. Hence, individual attitudes toward ambiguity are likely to influence who is drawn to and who remains in which decision environment. The actions of the two entrepreneurs may then fundamentally differ due to their different ambiguity attitudes even if their perceived chances of success, i.e., their beliefs about their respective opportunities, are identical. Importantly, the two entrepreneurs may differ not only in how much ambiguity they are willing to tolerate, but also in how willingly they adapt their actions when their beliefs about success shift. For the food truck owner, survival may depend on rapid adjustments: changing the menu, revising prices, or moving to a new location when receiving negative signals. By contrast, the deep-tech entrepreneur may often need to endure prolonged ambiguity and may need to signal a strong commitment to a single story to sustain investor and partner confidence. The current conceptualization of tolerance for ambiguity in entrepreneurship cannot capture these two differing aspects of decision-making; instead, it treats them as one concept.

---

[1] Ambiguity is a type of uncertainty where the potential events can be enumerated, but their probabilities are not objectively known. We will discuss the definition in detail in the following section.

This study advances our understanding by theoretically distinguishing and empirically measuring two components of ambiguity attitudes that shape entrepreneurial action. The first component, ambiguity aversion (or, its inverse, tolerance for ambiguity), refers to an individual's willingness to bear ambiguity; more ambiguity-tolerant individuals are more willing to take actions associated with greater levels of ambiguity. The second component, ambiguity sensitivity, reflects how individuals discriminate (or fail to discriminate: ambiguity insensitivity) between different levels of perceived chances of success.[2] Ambiguity-insensitive individuals treat different ambiguous opportunities similarly, regardless of whether they estimate the odds of success as high or low. In the extreme, a fully ambiguity-insensitive individual acts the same in an ambiguous situation, no matter how her subjective probabilities change. In contrast, those with high ambiguity sensitivity adjust their actions more closely in line with their subjective probability estimates. We suggest that these two components capture relevant and distinct dimensions of ambiguity attitudes that affect entrepreneurial actions.[3]

---

[2] The full description of ambiguity (in)sensitivity is "ambiguity-generated likelihood (in)sensitivity," often also called "a-(in)sensitivity" (Dimmock et al., 2016).

[3] Parts of the literature characterize ambiguity aversion as primarily a motivational factor and ambiguity sensitivity as primarily a cognitive factor (Dimmock et al., 2016; Baillon et al., 2021; Gao et al., 2024; Henkel, 2024). We do not prescribe the use of these interpretations, and these interpretations are not necessary for our theorizing, analysis, or conclusions.

Not all entrepreneurs have the same growth intentions (Davidsson, 1991; Wiklund et al, 2003; McKelvie et al., 2021). We propose that entrepreneurs with different growth intentions have different ambiguity attitudes. A small set of entrepreneurs want to grow fast, while most instead have intentions only to serve a smaller local market and remain small (Hurst and Pugsley, 2011). In the former category, we find scale-ups, those that target the most recent fad or technology, and these entrepreneurs are most likely to contribute to creative destruction (Schumpeter, 1911; Aghion et al., 2021). In the latter category are most self-employed lawyers, physicians, carpenters, restaurateurs, and shop owners, to name just a few of many small business owners. In this article, we distinguish these groups empirically through their choice of legal form: we label and measure them as entrepreneurs with incorporated businesses and the self-employed with unincorporated businesses, respectively.

Entrepreneurs with higher growth intentions typically choose to incorporate their business (Stiglitz and Weiss, 1981; Levine and Rubenstein, 2017) because this legal structure facilitates raising external equity and scaling rapidly. An entrepreneur with lower growth ambitions may instead choose a less expensive and administratively less burdensome legal form but then faces full downside loss liability. Incorporation is therefore more common in highly ambiguous environments, where ventures either scale rapidly or fail quickly (Haltiwanger et al., 2013). To secure external capital and maintain legitimacy, growth-oriented entrepreneurs often need to remain committed even when early signals are unfavorable (Garud et al., 2014). By contrast, self-employed individuals tend to operate in more predictable markets, but as they face full downside liability, their success and survival is more closely tied to their ability to respond quickly to market signals. Because of these differences, we expect incorporated entrepreneurs to exhibit lower ambiguity aversion, meaning a greater willingness to operate under high ambiguity. Instead, self-employed individuals should exhibit higher ambiguity sensitivity, allowing them to quickly adapt to market changes, while incorporated entrepreneurs may be less sensitive to new information in their decision making, maintaining their strategic narrative and persisting with original plans despite contrary signals.

Empirically, we compare ambiguity attitudes between self-employed individuals, incorporated entrepreneurs, and employees using incentivized experimental tasks administered to a large Dutch population sample across six waves (2018-2021). Consistent with our expectations, our findings reveal that ambiguity aversion is negatively associated with being an incorporated entrepreneur relative to being an employee. Ambiguity sensitivity is positively associated with being self-employed but not with incorporated entrepreneurship relative to employment. These results hold even after controlling for risk preferences, optimism, cognitive skill, and demographics, providing new evidence that attitudes toward ambiguity explain entrepreneurial behavior beyond risk aversion. The results also remain robust to an alternative operationalization of growth orientation.

Our findings imply that different types of entrepreneurs differ not only in their willingness to bear ambiguity but also in how they translate updated beliefs into action. This distinction helps explain why some entrepreneurs persist in a chosen course despite new information, while others pivot more readily. By showing that these two components of ambiguity attitudes shape entrepreneurial behavior in distinct ways, we offer a novel decision-theoretic perspective on how ambiguity attitudes influence entrepreneurial action. This perspective opens new avenues for research and practice, for example concerning the design and targeting of entrepreneurship training and support programs.

## 2. Literature review

### 2.1 Ambiguity, ignorance, and risk

Knightian uncertainty is central to canonical theoretical approaches in entrepreneurship, including the judgment-based view (Foss and Klein, 2012), entrepreneurial action theory (McMullen and Shepherd, 2006), and the theory-based view of entrepreneurship (Felin and Zenger, 2009; Camuffo et al., 2024; Ehrig and Zenger, 2024). However, despite its importance, there is no consensus on the nature of Knightian uncertainty, and multiple perspectives exist in the literature (for recent reviews, see Townsend et al., 2024, and Dorobat et al., 2026). Some scholars emphasize the absence of basis for classifying instances into groups (Langlois and Cosgel, 1993), the impossibility of forming insurance contracts (LeRoy and Singell, 1987), or the "inability to articulate and communicate, or transfer, estimates about the future" (Foss and Klein, 2012, p. 85). Others focus on the subjective experience of uncertainty, describing it as a sense of doubt that inhibits or delays action (McMullen and Shepherd, 2006; Packard et al., 2021). Some authors conceptualize Knightian uncertainty as unknowability, a universal epistemological limit (Ramoglou, 2021; Arend, 2022), while others argue that Knightian uncertainty comprises both mitigable epistemic uncertainty and immitigable aleatory uncertainty rooted in genuine indeterminism (Packard and Clark, 2020; see also Townsend et al., 2024).

A common element across most perspectives is that under Knightian uncertainty, decision-makers lack access to objective probabilities. Indeed, Knight (1921) distinguished between three types of decision situations based on the availability and nature of probability information. In the first situation, a priori objective probabilities are available to the decision maker. In the second situation, empirical evidence allows the decision maker to form statistical probabilities about the realization of an event. For instance, an insurance company can assess the probability of a car accident by analyzing historical data of similar cases. In the third situation, there is little or no basis for comparison. The decision maker cannot establish the probability of an event and must instead make a decision based on a subjective estimate. Knight argued that this is the type of uncertainty that characterizes entrepreneurial decision-making.

Parker (2025) offers a useful classification that bridges the Knightian uncertainty literature and formal decision theory by distinguishing *Type I uncertainty*, where the potential states of the world (events) can be enumerated but their probabilities are unknown, from *Type II uncertainty*, where some possible states of the world are themselves unknown and potentially unimaginable, making it impossible to form an exhaustive set of probability judgments. Both types fall within the scope of Knightian uncertainty but represent different degrees of the knowledge problem entrepreneurs face, not different interpretations of the concept (see also Packard et al., 2017, for a related typology). Type I uncertainty corresponds to what decision theorists and economists often call *ambiguity* (e.g., Ellsberg, 1961; Nishimura and Ozaki, 2004;

Brenner and Izhakian, 2022). An example is an urn known to contain red and black balls (and nothing else), so only red or black balls can be drawn from it, but the numbers of red and black balls are unknown. Thus, the probabilities of drawing either color from the urn are unknown, but individuals can form subjective beliefs about these probabilities. Another example of ambiguity is the direction the Dow Jones stock market index will move over the next 100 days: the two possible events (up or down) are known, but the probabilities are unknown. Ambiguity contrasts with Type II uncertainty, which Dorobat et al. (2026) refer to as *ignorance*, where it is not even possible to enumerate the potential states of the world. In the example of the urn, it is unknown how many different colors of balls the urn contains or if it even contains something entirely different. As the possibilities are infinite, and some may not even be imaginable, one cannot assign subjective probabilities to all possible events. In contrast to both ambiguity and ignorance, *risk* is a situation where both the potential events and their associated probabilities are objectively known. In case of the urn, it may for example be known that the urn contains five black balls and five red balls, so the probability of drawing a black ball is objectively 50%.[4]

In this paper, we study how entrepreneurs make decisions under ambiguity, i.e. Type I Knightian uncertainty. We do not claim that ambiguity represents the whole scope of uncertainty that entrepreneurs face. Entrepreneurs likely encounter Type II uncertainty as well, particularly in the earliest stages of venture creation (Packard et al., 2017). However, we argue that studying entrepreneurial decision-making under ambiguity is relevant and offers significant conceptual and empirical value. First, many entrepreneurial decisions are not made under conditions of absolute ignorance (Parker, 2025; Dorobat et al., 2026). Knight (1921, p. 199) noted that many entrepreneurial decisions can be characterized as involving "action according to opinion, of greater or less foundation and value, neither entire ignorance nor complete and perfect information, but partial knowledge", making ambiguity a useful proxy.

---

[4] Although some frameworks arrange risk and uncertainty on a single continuum (Dequech, 2011), our focus follows decision science in treating preferences over known vs. unknown probabilities as qualitatively distinct.

Second, a closely related view holds that entrepreneurs make decisions based on *subjective* uncertainty rather than objective uncertainty (Packard et al., 2021). Even under conditions of ignorance, entrepreneurs may perceive a situation differently based on their knowledge, experience, and cognitive frameworks (e.g., Grégoire and Shepherd, 2012). This theoretical position is supported by recent empirical findings. Angus et al. (2023) found that entrepreneurial actions were significantly related to perceived uncertainty but not to objective unpredictability. They explain: "while the future may be unpredictable to varying degrees, the mind can always imagine possible scenarios through mental simulation processes … resulting in the prediction and expectation of possible outcomes given specific inputs, including possible actions" (Angus et al., 2023, p. 1152). Similarly, Townsend et al. (2024, p. 465) observe that "entrepreneurs can strongly believe a future state of a decision environment will emerge even when existing evidence is weak and/or the environment is characterized by a high degree of indeterminism." The absence of precise information does not prevent entrepreneurs from forming beliefs about their chances of success; these beliefs may be poorly calibrated (e.g., Hayward et al., 2006; Kraft et al., 2022; Chochoiek et al., 2025), but they still guide action. When entrepreneurs operate with such subjective beliefs in the absence of known probabilities, they face ambiguity, and it is under these conditions that ambiguity attitudes become critical.

Third, entrepreneurs have agency in transitioning from ignorance to a state of ambiguity. For example, Packard et al. (2017) model this as a recursive process: entrepreneurs iteratively narrow the set of possible outcomes and actions, whether through effectuation techniques (Sarasvathy, 2001) or causal reasoning. The theory-based view of entrepreneurship proposes that entrepreneurs can systematically transition from ignorance to manageable ambiguity through theorizing and experimentation (Zellweger and Zenger, 2023; Camuffo et al., 2024; Ehrig and Zenger, 2024). Because entrepreneurial processes are inherently iterative (McMullen & Dimov, 2013; McMullen, 2015), entrepreneurs repeatedly make such transitions from ignorance to states of ambiguity, and consequently to action and learning. Therefore, studying ambiguity attitudes addresses a recurrent and important aspect of entrepreneurial decision-making.

To sum up, while understanding how an entrepreneur navigates deeper forms of uncertainty such as ignorance remains critically important, we argue that studying entrepreneurial decision-making under ambiguity addresses a relevant feature of the entrepreneurial process. Studying ambiguity offers a tractable framework for empirical research, enabling precise measurement and quantification of attitudes that is, to our knowledge, not possible when studying situations where decision-makers cannot enumerate possible states of the world (for a discussion of the intractability of Type II uncertainty, see Parker, 2025). This focus also answers recent calls to conduct micro-level studies that investigate the question: "how do entrepreneurs cope with KU [Knightian Uncertainty] as ambiguity" (Dorobat et al., 2026, p. 212).

*2.2 Ambiguity attitudes in entrepreneurship*

Attitudes toward ambiguity reflect the propensity to act in ambiguous situations based on beliefs, whether biased or unbiased, about the chances of success. This section reviews different ideas and perspectives on how such attitudes have been theorized and measured in the entrepreneurship literature.

A foundational perspective on these attitudes comes from Frenkel-Brunswick (1951), who proposed the concept of "tolerance for ambiguity" as a stable and broad predictive variable across behavioral settings. She suggested that individuals with high tolerance for ambiguity perceive ambiguous situations as desirable, challenging, and interesting. Tolerance for ambiguity should be particularly relevant for entrepreneurs who often face situations with unknown probabilities of success. Sexton and Bowman (1985) first applied Frenkel-Brunswick's concept to entrepreneurship, comparing entrepreneurship students with students in other fields, and finding that entrepreneurship students exhibited greater tolerance for ambiguity. The concept has also shaped research on entrepreneurial orientation, which originally characterized the construct's attitudinal dimension as "attitude toward risk," before recent work clarifies that it encompasses attitudes toward ambiguity where probabilities are unknown (Anderson et al., 2015; Putniņš and Sauka, 2020). However, no clear consensus has emerged from empirical research, with several studies finding no significant relationship between tolerance for ambiguity and entrepreneurial intentions or the decision to become an entrepreneur (see Gurel et al., 2010; Altinay et al., 2012; Furnham and Marks, 2013). These inconsistent findings may stem from fundamental conceptual and measurement challenges. As Furnham and Marks (2013, p. 718) note, as of 2013, "...despite work on these subtly different and related concepts there is still no very clear operational definition of tolerance for ambiguity at the facet level or a clear differentiation between the manifestations and correlates of tolerance for ambiguity."

Researchers have sought to address these challenges using a measurement approach developed in behavioral economics. This approach measures tolerance for ambiguity through revealed preferences rather than self-reported attitudes. For instance, instead of asking subjects for their perceptions about their behavior, such as asking for the extent to which they would tolerate an ambiguous situation on a scale from 1 to 7, the approach directly measures their behavior when facing ambiguous outcomes. A common approach to measure tolerance for ambiguity in behavioral economics is to test decision makers' choices between drawing from an urn with an unknown composition of balls and an urn with a known composition of balls. A preference for the unknown urn is interpreted as a higher tolerance for ambiguity, while a preference for the known urn suggests ambiguity aversion (see Ellsberg, 1961, and further discussion to follow). Using this method, Holm et al. (2013) and Koudstaal et al. (2016) found no systematic differences in tolerance for ambiguity between entrepreneurs and non-entrepreneurs. Yet, the Ellsberg task has been criticized as artificial and potentially not very informative for real-life decisions (Camerer and Weber, 1992; Li et al., 2018), and this may explain why no relationship has been found with entrepreneurship.

Beyond individual-level studies, another stream of research explores the relationship between entrepreneurship and country-level uncertainty avoidance, also called intolerance of ambiguity—the extent to which the members of a culture feel threatened by ambiguous or unknown situations (Hofstede et al., 2010, p. 191). Several studies suggest that low uncertainty avoidance is associated with entrepreneurial orientation (Mueller and Thomas, 2001), entrepreneurial entry (McGrath et al., 1992; Autio et al., 2013), rates of innovation (Shane, 1993), entrepreneurial risk-taking and proactivity (Kreiser et al., 2010), and the use of entrepreneurial orientation rhetoric (Watson et al., 2019). However, evidence is mixed, with some studies finding a negative relationship between uncertainty avoidance and high-growth entrepreneurship (Bowen and De Clercq, 2008), no relationship with growth practices (Autio et al., 2013), or even a positive relationship with business ownership (Wennekers et al., 2007). Most of these studies rely on versions of Hofstede's survey on cultural values (Hofstede et al., 2010). While informative, this measure cannot differentiate between attitudes toward risk and ambiguity, and Hofstede himself emphasizes that it is a societal-level rather than an individual-level measure and therefore cannot be used to differentiate between members of the same culture.

In light of these challenges, it becomes essential to define and measure ambiguity attitudes precisely and to distinguish them from other related constructs such as risk attitudes, because ambiguity attitudes have unique theoretical implications. For instance, traditional risk-based models, in which probabilities of outcomes are known, predict that an increase in uncertainty (viewed as a higher variance in returns) leads prospective entrepreneurs to search longer for higher returns opportunities. However, when facing ambiguity about the distribution of returns itself, theory predicts the opposite: entrepreneurs will search less and accept lower-value opportunities earlier (Nishimura and Ozaki, 2004). These markedly different behavioral responses highlight why developing precise concepts and measurements of ambiguity attitudes—differentiated from risk attitudes—is crucial for understanding entrepreneurial actions.

We suggest that two factors may help explain why prior findings on an entrepreneur's attitudes toward ambiguity have remained inconclusive. First, prior work has largely conceptualized ambiguity attitudes as a single trait. Recent advances in decision theory, however, suggest that ambiguity attitudes consist of at least two distinct dimensions (e.g., Dimmock et al., 2016; Baillon et al., 2018, 2021): (1) ambiguity aversion, reflecting willingness to act when probabilities are unknown, and (2) ambiguity sensitivity, reflecting how strongly actions adjust to changes in subjective probability assessments. Second, research on ambiguity attitudes has generally treated entrepreneurs as a homogeneous group. Yet, self-employed individuals and incorporated entrepreneurs differ in ways that could be associated with how they act under ambiguity. The mixed results in prior empirical analysis may therefore be partially due to pooling these two types of entrepreneurs together. We propose to address both issues by separating the two dimensions of ambiguity attitudes and by analyzing the self-employed and incorporated entrepreneurs separately.

## 3. Theory: ambiguity aversion and ambiguity sensitivity

The theoretical framework that we use distinguishes between two conceptually and behaviorally distinct dimensions of ambiguity attitudes: ambiguity aversion and ambiguity sensitivity.

### 3.1 Ambiguity aversion

In a seminal paper, Ellsberg (1961) illustrated the concept of ambiguity aversion through two thought experiments involving urns with known and unknown compositions. Consider a decision-maker facing two urns, each containing 100 balls that can be either red or black. The first urn (urn K) has a known composition of 50 red balls and 50 black balls. The second urn (urn U) also contains 100 balls, but the exact composition of red and black balls is unknown. When the decision-maker is asked to bet on the color of a ball drawn from urn U, offering $100 if the chosen color is drawn, she typically shows indifference between the colors, suggesting she assigns a subjective probability of 50% to drawing either color. However, when given a choice between a bet that pays $100 if a black ball is drawn from urn K and a similar bet on urn U, most people prefer to bet on the urn with the known composition. This preference holds regardless of the color specified for the bet and is therefore not due to different beliefs about the likelihood of drawing a black or red ball from urn U. This illustrates the concept of ambiguity aversion: a preference for events with known probabilities over those with unknown probabilities. This concept differs from risk aversion, which reflects a preference for certain over risky outcomes. For example, a risk-averse individual would prefer a guaranteed $50 over a 50% chance of winning $100, even though the expected value is the same. Instead, an ambiguity-averse individual would prefer a 50% chance of winning $100 over an option that she believes, but does not know for sure, has a 50% chance of winning $100.

Importantly, ambiguity aversion does not capture differences in subjective probabilities, which remain equal for both urns, but how people act on their beliefs under ambiguity. In the case of Ellsberg's urn experiment, ambiguity aversion is measured by comparing the valuation of bets, or willingness to pay (WTP) for bets, which involve risk versus ambiguity. Thereby, ambiguity aversion captures the preference for reducing the level of ambiguity.

*3.2 Ambiguity sensitivity*

Ambiguity sensitivity captures the decision makers' capacity to discriminate between varying levels of perceived probabilities (Baillon et al., 2018). When ambiguity sensitivity is low, people treat events with different perceived likelihoods of success similarly "as one blur" (Baillon et al., 2021, p. 6) and insufficiently respond to changes in perceived likelihoods (Dimmock et al., 2016; Baillon et al., 2018). Consequently, when ambiguity sensitivity is low, WTP for a bet is similar regardless of the person's beliefs about the chances of success. The influence of beliefs on actions is therefore attenuated.

Importantly, ambiguity sensitivity does not capture differences in subjective probabilities themselves but rather the differences in how people translate those probabilities into behavior, i.e., the responsiveness of the WTP to changes in perceived likelihoods. At the extreme, a person with complete ambiguity insensitivity would choose the same action in an ambiguous situation regardless of the subjective probabilities of success.

*3.3 The difference between ambiguity aversion and ambiguity sensitivity*

To illustrate the difference between ambiguity aversion and ambiguity sensitivity, consider two investors, A and B, evaluating an entrepreneurial investment project with ambiguous chances of success. In case of success, the project generates a return of $1 million to the investor, but otherwise the investment is lost. Both investors subjectively perceive the likelihood of success as 40%. Further assume that both investors have identical risk attitudes and identical access to resources. This implies that, if the 40% success

probability was objectively known, like in a roulette gamble, both would invest the same maximum amount (for instance $400k). However, given that the investment is ambiguous, Investor A is willing to invest a maximum of $300k, whereas Investor B would invest only $200k. The difference cannot be attributed to risk preferences as both investors have the same degree of risk aversion, and the perceived likelihood of success is the same. Because Investor B exhibits a stronger reluctance to bet when probabilities are subjectively perceived, she is more ambiguity averse than Investor A.

**Figure 1: Illustration of the difference between ambiguity aversion and ambiguity sensitivity**

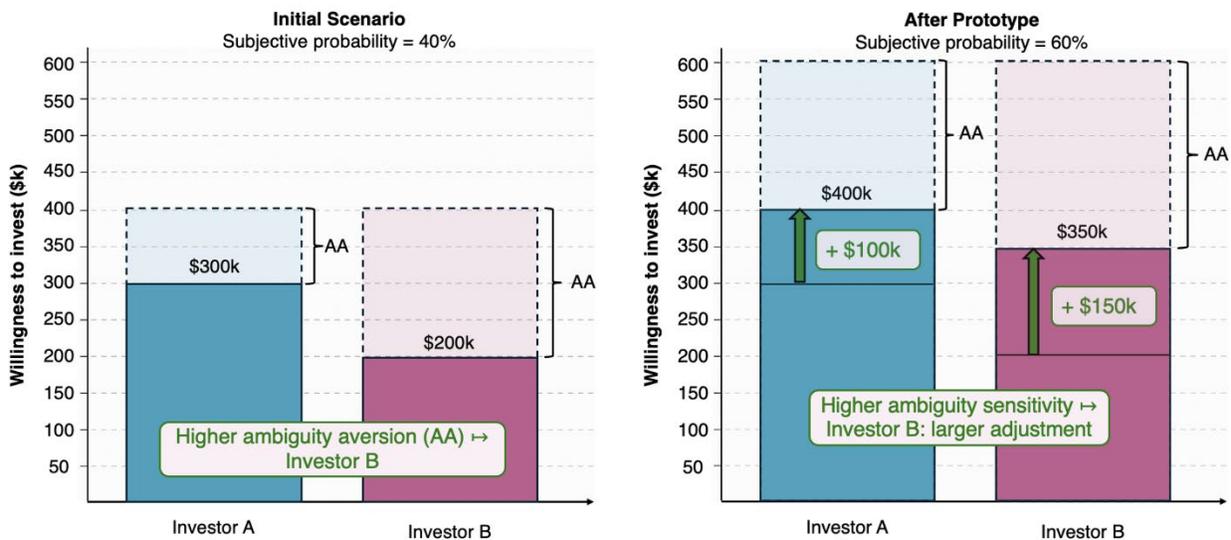

*Note*: Investors A and B have identical risk attitudes and beliefs. For simplicity, we assume risk neutrality; that is, if probabilities were objective and known, both investors' willingness to invest would be $400k in the initial scenario and $600k after prototyping (as represented by the dashed boxes).
Ambiguity Aversion (AA): Investor B is more ambiguity averse than investor A as she consistently invests less than A when probabilities are unknown (lower tolerance for ambiguity).
Ambiguity Sensitivity: Investor B is more sensitive to ambiguity than investor A as she shows a larger adjustment to changes in beliefs (+$150k vs. +$100k).

Now suppose that a prototype is built and shows promising results. Given this new information, both investors update their beliefs about the project's likelihood of success, increasing their subjective probability estimates from 40% to 60%. Investor A is now willing to invest $400k instead of $300k, while Investor B is willing to invest $350k instead of $200k. These new investment choices confirm that Investor B is more ambiguity averse than Investor A, as she is still not willing to invest as much as Investor A despite having the same updated beliefs. In addition, Investor B increases her willingness to invest by $150k while Investor A increases hers by only $100k in response to the same change in perceived probability. This indicates that Investor B is more sensitive to changes in perceived probabilities than Investor A, i.e., she exhibits higher ambiguity sensitivity. This difference in their behavior is not due to differences in beliefs, as the subjective probability of success increased from 40% to 60% for both investors, and it is also not due to different risk attitudes. Rather, it is due to a difference in ambiguity sensitivity, which captures the degree to which decision makers adjust their actions when their subjective probabilities change. Figure 1 illustrates this example.[5]

*3.4 Heterogeneity across entrepreneurial contexts*

We distinguish between two archetypical entrepreneurs: those intending to grow their business rapidly and those who do not.[6] We argue that these two categories differ fundamentally in the nature and degree of ambiguity that they face. Accordingly, we claim that their attitudes toward ambiguity are likely to differ, warranting separate analyses.

Some entrepreneurs tend to have an interest in growing their businesses rapidly (Davidsson, 1991; Wiklund et al, 2003; McKelvie et al., 2021), but far from all. Most entrepreneurs instead have little to no intention to grow their business and are comfortable keeping their business small (Hurst and Pugsley, 2011). For those with ambitious growth intention, rapid growth is however not ordained, and many fail. Such an entrepreneur is therefore exposed to considerable variability in outcomes. The typical growth path for these ventures is characterized as up-or-out (Haltiwanger et al., 2013). Failing quickly is often associated with unexpected events, such as a technology failing to meet market needs or falling short of the performance levels initially expected. For example, some entrepreneurs are exposed to large amounts of ambiguity because they choose to start ventures in emerging sectors such as new battery technologies or clean energy. In these sectors there is currently high ambiguity about which products or business models will prove commercially valuable. In the future, other sectors will instead temporarily experience this state of high ambiguity.

---

[5] Note that this example and Figure 1 illustrate 'local' ambiguity aversion and sensitivity at specific subjective probability levels (40% and 60%). In contrast, our empirical measures capture a 'global' assessment of both parameters across the full spectrum of likelihoods, aggregating behavior over multiple subjective probability levels rather than focusing on specific points.

[6] This is not to say that there are no other ways one might classify different types of entrepreneurs. For example, some individuals unable to find paid employment may become entrepreneurs out of necessity (O'Donnell et al., 2024), whereas others may be described as opportunity-based (Fairlie and Fossen, 2020). In this article, we focus on differences between entrepreneurs that are associated with differences in ambiguity attitudes.

Conversely, a second and much larger group of entrepreneurs typically show less intention to innovate and grow (Stewart and Roth, 2007; Hurst and Pugsley, 2011). Most of these intend to provide an existing service to an existing market. They also exhibit much smaller variations in objective measures of growth (Coad et al., 2014; Åstebro and Tåg, 2017; Levine and Rubinstein, 2017; Botelho et al., 2021; Coad and Karlsson, 2022). They tend to engage in locally bounded and established activities where information about local demand, prices, and competitive offerings is more readily observable, such as food services or performing home repairs (Hurst and Pugsley, 2011). Consequently, such an entrepreneur operates in a relatively signal-rich environment and is generally less exposed to extreme ambiguity than a growth-oriented entrepreneur pursuing novel opportunities.

One way to distinguish these two groups empirically is to use whether their business is incorporated or not. Incorporation provides limited liability, protecting the owner's personal assets from business losses and encouraging greater risk-taking in business decisions (Stiglitz and Weiss, 1981; Levine and Rubinstein, 2017). Unincorporated self-employed, on the other hand, have full personal liability to pay back claimholders in case of business failure. Entrepreneurs who intend to grow often choose incorporation because it facilitates raising equity and sharing both the upside opportunity and risk with investors (Levine and Rubenstein, 2017). By contrast, remaining unincorporated is cheaper and administratively simpler when entrepreneurs have no intention to raise venture capital or pursue rapid expansion. In our main analysis, we therefore use legal form as an empirical proxy for growth orientation, and for ease of understanding and compactness in presentation we call the two latent classes self-employed and incorporated entrepreneurs. To ensure robustness, we also test realized firm growth as an alternative indicator of growth orientation.

The prior discussion suggests that a growth-oriented entrepreneur is exposed to considerable ambiguity that she must tolerate, while those who cannot tolerate such ambiguity pursue entrepreneurial paths with lower ambiguity or stay employed. Hence, we propose that these two types of entrepreneurs, incorporated entrepreneurs and self-employed, are associated with fundamentally different attitudes toward ambiguous probabilities of success. Specifically, we propose that incorporated entrepreneurs tend to have a higher tolerance for ambiguity as compared to employees, spurring them to take on situations with high ambiguity.[7]

An entrepreneur with high growth intentions is more likely to try and raise external equity capital to finance rapid scale-up. Raising equity involves convincing other parties to believe in your plans and forecasts. An important part of creating that confidence is good storytelling and persuasion skills (Chen et al., 2009; Garud et al., 2014; Clough et al., 2019; Hu and Ma, 2025; Stefanidis et al., 2025). Because of this need for providing a coherent story, an incorporated entrepreneur is less likely to be ambiguity sensitive: she is less likely to change her course of action when adjusting her subjective probabilities. The environment for resource accumulation in these ventures demands lower sensitivity to ambiguity.

In contrast, the self-employed, lacking limited liability, are fully exposed to downside losses (Stiglitz and Weiss, 1981; Levine and Rubinstein, 2017). Because of this direct personal exposure, their financial survival depends on quickly detecting and responding to changes in local market conditions. Such environments should attract individuals capable of quick reactions to changing conditions. The contexts characterized by abundance of market signals enable and reinforce this pattern. Consequently, we propose that the self-employed are more likely to be ambiguity sensitive than employees: they act more sharply on changes in their beliefs.

---

[7] This expectation is consistent with Bonilla and Cubillos (2021, p. 63), who theoretically predict that "increases in ambiguity aversion reduce entrepreneurial activities".

## 4. Empirical methodology

### 4.1 Challenges of studying ambiguity attitudes in natural contexts

A central empirical challenge in studying ambiguity attitudes is separating decision makers' attitudes from their beliefs. For example, an entrepreneur's decision to turn down a $1 million buyout offer is driven both by the entrepreneur's beliefs about the venture's value and by her attitudes toward the ambiguity surrounding the decision. In the illustrative example used in Section 3.3, we held investors' beliefs constant. In practice, however, beliefs are often inaccessible to researchers, complicating the task of separating their influence from that of attitudes.

This identification problem explains why much prior work has used 'artificial' sources of uncertainty, often drawing on variations of Ellsberg's thought experiment. In such designs, researchers can 'control' beliefs, for instance by assuming a 50% chance of drawing a red ball from an urn with an unknown composition. A drawback of such experiments is that they may not be very informative about ambiguity attitudes in real-world contexts such as entrepreneurship because they are unrealistic and decontextualized.

Recent advances in research on decision-making under uncertainty have introduced tools that allow the measurement of ambiguity attitudes in natural contexts, free from distortions due to beliefs. One approach, the events-exchangeability method (Abdellaoui et al., 2011), requires measuring beliefs first before assessing ambiguity attitudes. A second approach, the belief-hedging method (Baillon et al., 2018), does not measure beliefs separately but instead neutralizes their effect.[8]

---

[8] We note that early literature has utilized judged probabilities to measure beliefs, exemplified by Tversky and Fox's (1995) two-stage model. However, this method is less suitable for distinguishing between beliefs and ambiguity attitudes, as these judged probabilities may also reflect aspects of ambiguity attitudes. For a discussion on this topic, see Abdellaoui et al. (2024).

While the events-exchangeability method has been used in the management literature to analyze decisions such as market entry or investment decisions (Gutierrez et al., 2020; Sonsino et al., 2022), it is complex due to the need to separately measure beliefs and risk attitudes (Abdellaoui et al., 2011). Instead, the belief-hedging method (Baillon et al., 2018) is easier to implement and more suitable for large-scale surveys. It relies on bets on complementary events to gauge ambiguity attitudes. For brevity, we relegate the technical details of the incentivized measurement procedure to Online Appendix A, where we first explain the intuition underpinning the method and then detail the structural model estimations used to measure the two components of ambiguity attitudes. In our empirical analysis, we will use our main measures of ambiguity attitudes derived from the belief-hedging method as well as a measure derived from an Ellsberg urns experiment to compare the predictive power of these alternative approaches.

*4.2 Data: randomly drawn population panel*

We use the Longitudinal Internet Studies for the Social Sciences (LISS) panel (Scherpenzeel, 2011), a probability sample created by randomly choosing postal addresses from the Dutch population register in collaboration with Statistics Netherlands.[9] Since 2008, participants have regularly completed online questionnaires. Every year, a fixed set of recurring core questionnaires is assigned to the participants, including the annual "Work and Schooling" survey, which captures employment and entrepreneurship. Researchers can also create their own questionnaires for the full panel or a subsample, and the resulting data are made available to researchers worldwide. Von Gaudecker et al. (2022) included incentivized tasks in six waves between 2018 and 2021 to elicit ambiguity attitudes in the context of the stock market using the belief-hedging method mentioned above and described in detail in Online Appendix A. These ambiguity attitudes are our key independent variables. The LISS has been used in entrepreneurship research (e.g., Hessels et al., 2014; Block and Petty, 2023), but this paper is the first to use the ambiguity module in the context of entrepreneurship. The source of uncertainty that we use (the future value of a financial asset) shares important features with the uncertainty entrepreneurs often face; in both cases, the decision maker must form beliefs about future outcomes whose underlying probability distribution is unknown and not knowable.

---

[9] If a selected household does not have a broadband internet connection or a computer, the required equipment is loaned to enable the household to take part in the panel. The random population sample provides an advantage (by avoiding self-selection) in comparison to surveys that allow people with internet access to sign up themselves to a pool of potential participants (as in, for example, MTurk or Prolific).

As motivated above, we distinguish between two types of entrepreneurs, the self-employed versus entrepreneurs running incorporated businesses. Our variable "self-employed" comprises survey respondents who indicate that they are self-employed, freelancers, or independent professionals. Our variable "incorporated entrepreneur" captures respondents who answered that they are directors of a limited liability or private limited company or a majority shareholder director of such. Both types of entrepreneurs may or may not have employees. There can be multiple ways to empirically distinguish between entrepreneurs aiming to rapidly grow their business companies versus entrepreneurs who are less ambitious. We use realized firm growth as an alternative measure for robustness, and a third measure of firm size for comparison. The firm growth distinction considers entrepreneurs with a growing number of employees supervised (on average over our observation period from 2018 to 2021) versus those with a stagnant or shrinking number of employees. A disadvantage of this measure is that the actual growth partially depends on the current success of the business, which may reflect external factors in addition to the type of entrepreneur. Our third distinction is between employers, who supervise at least one employee, versus non-employers. This measure is less preferred by us as it partially reflects the minimum efficient size of a business; for example, a restaurant usually cannot run without employees, although the owner may not be growth oriented, whereas a software consultant may start alone despite being growth oriented. The three categorizations of entrepreneurs overlap to a large extent. For example, among the incorporated entrepreneurs, 67% are also employers, and among those who report themselves to be self-employed, 76% are also non-employers (self-employment does not necessarily imply working without employees).

In our main and preferred analysis, we restrict the sample to working-age individuals above 20 years and below 65 years of age. We classify individuals based on their status contemporaneously to the ambiguity attitudes measurement between 2018 and 2021. We classify an individual as an incorporated entrepreneur if the individual was observed as an incorporated entrepreneur in at least one of the annual "Work and Schooling" interviews during this four-year period. Among the remaining individuals, we classify those as self-employed who were observed as self-employed at least once during this period but never as incorporated entrepreneurs. We label the remaining individuals as employees if they were observed as an employee in permanent or temporary employment, as an on-call employee, or as a temp staffer at least once during this period. Those who were never observed working between 2018 and 2021 are excluded from the sample as we focus on differences among the working population.

The LISS includes a significant number of individuals of retirement age, who are excluded from our main estimations based on their current occupational status. In a robustness check, we include these older individuals; to do so, we rely on the previous employment for individuals no longer working. Specifically, we add individuals aged 65 years and older to the sample, including retired individuals, while we still exclude individuals aged 20 or younger, and we base the classification of individuals into the occupational groups on their current and previous employment as observed in the annual "Work and Schooling" questionnaires in the full LISS panel going back as far as 2008 for some individuals. In this extended sample, we classify individuals as incorporated entrepreneurs if they were ever observed as an incorporated entrepreneur, as self-employed if they were ever observed as self-employed but never as an incorporated entrepreneur, and as employees if they were ever observed as an employee but never as an incorporated entrepreneur or self-employed, and we exclude individuals from the sample if they were never observed working in the entire LISS panel. The advantage of this alternative approach is that we can assess whether our main results hold across a broader sample of individuals. Results appear robust across the two alternate samples.

*4.4 Control variables*

In our econometric estimations, we control for cognitive and preference variables as well as demographics that may be correlated with ambiguity attitudes and entrepreneurial choices.

We intend to estimate the unique contributions of ambiguity attitudes as determinants of entrepreneurship in addition to and independent of conventional risk attitudes. Therefore, as a control variable, we generate a *risk aversion* index based on measurements in the LISS in the November 2018 and November 2020 surveys. We use the experimentally validated preference survey module developed by Falk et al. (2023), including a qualitative component (a question for the general willingness to take risks answered on a 10-point scale) and a quantitative component that is based on elicited certainty equivalents for five risky lotteries (for details, see Von Gaudecker et al., 2022). After standardizing each component, the qualitative component receives a weight of 53%, as suggested by Falk et al. (2023). We take the mean of the 2018 and 2020 measurements for each individual.

*Optimism,* which reflects a general tendency to hold positive beliefs about future outcomes, may influence entrepreneurial choices (Åstebro et al., 2014). We control for optimism using the Revised Life Orientation Test by Scheier et al. (1994), which is included in the annual Personality questionnaire of the LISS. This inventory consists of six items, such as "I'm always optimistic about my future" or "In uncertain times, I usually expect the best", as well as four filler items. Respondents evaluate how much the statements apply to them on a 5-point scale. We use the first observed measurement for each individual during the 2018-2021 period (when the ambiguity attitudes were elicited). We average over the 6 items; some of the items are reversed to achieve that higher values always indicate more optimism.

Ambiguity attitudes, especially ambiguity sensitivity, which has sometimes been characterized as a cognitive component (Dimmock et al., 2016; Baillon et al., 2021; Gao et al., 2024; Henkel, 2024), may be correlated with cognitive skill; controlling for this allows us to isolate the specific effects of ambiguity attitudes. *Cognitive skill* was measured in the LISS within the ambiguity module. The index is based on financial numeracy, probabilistic numeracy (both elicited in November 2018 and November 2020), and basic numeracy (elicited in January 2019 and November 2020). Controlling these particular aspects of cognitive skills is potentially important because the measurement of ambiguity attitudes uses the context of the stock market. The financial numeracy component consists of four questions that are a subset of the questions suggested by Van Rooij et al. (2011) and tests the ability to deal with interest rates and inflation. The probabilistic numeracy component uses five questions proposed by Hudomiet et al. (2018) and two questions added by Von Gaudecker et al. (2022) to test the understanding of probabilities. The basic numeracy component has been used before, for example, in the English Longitudinal Study of Ageing (Steptoe et al., 2013). It asks four or five questions involving numerical calculations in the contexts of money and health; the first three questions are the same for everybody, and the following questions are adjusted based on the correctness of the first questions. Von Gaudecker et al. (2022) provide the questionnaire wordings. For each component, we count the number of correct answers, standardize, and take the mean over the two measurements for each individual. The cognitive skill index gives equal weight to each component.

In terms of demographics, we control for age, gender, marital status, and the number of children in the household. Educational attainment is included using two indicator variables for upper secondary and tertiary education; below upper secondary education is the omitted base category.

*4.5 Econometric method*

We estimate ambiguity aversion and ambiguity sensitivity for each individual by maximum likelihood in a random-utility model (see Train, 2009; Von Gaudecker et al., 2022). These are our main independent variables. The technical procedure that we follow is well established in the literature (see Baillon et al., 2018; 2021) and is described in Online Appendix A.

To analyze the relationship between ambiguity attitudes and the two types of entrepreneurs while controlling for other relevant factors we estimate Probit models. We use one cross-section of data, as we have one estimated value of the two ambiguity parameters for each individual. The Probit models are based on the latent index equation:

$$y_i = \beta_0 + \beta_1 \, ambiguity\_aversion_i + \beta_2 \, ambiguity\_sensitivity_i + \gamma X_i + \varepsilon_i.$$

In the two Probit models, the dependent variable, $y_i$, is a binary variable indicating either self-employment or incorporated entrepreneur, respectively. We include both ambiguity attitudes—ambiguity aversion and ambiguity sensitivity—simultaneously because we are interested in the unique effect of each attitude, holding the other constant. The two attitudes are correlated (see Table 2), so omitting one factor would be expected to lead to a bias in estimating the coefficient of the other factor.[10] The vector $X_i$ includes the set of control variables discussed above. We report average marginal effects (or average effects of discrete changes from 0 to 1 in the case of indicator variables) with robust standard errors.

## 5. Empirical results

### 5.1 Descriptive statistics

Table 1 provides descriptive statistics by employment category where we distinguish between employees, the self-employed, and entrepreneurs with incorporated businesses. The upper panel shows the sample of working-age individuals segmented by their current employment, and the lower panel shows the extended sample including older individuals segmented by their current and previous employment. The preference and cognitive variables (ambiguity aversion, ambiguity sensitivity, risk aversion, optimism, and cognitive skill) are standardized based on the extended sample such that they have mean zero and unit standard deviation in the combined sample shown in the lower panel with 1,782 observations.

[ Insert Table 1 about here. ]

---

[10] We also estimated models omitting ambiguity aversion or ambiguity sensitivity, respectively. The results remained similar, indicating that the omitted variable bias is not large when only including one of the two variables.

Consistently across the two samples, incorporated entrepreneurs have a lower level of ambiguity aversion on average than the other two groups, and both the self-employed and incorporated entrepreneurs exhibit higher ambiguity sensitivity than employees. Both the self-employed and incorporated entrepreneurs are less risk averse than employees, confirming results from the literature, and incorporated entrepreneurs are the most optimistic. Incorporated entrepreneurs also score best on the cognitive skill test. This is consistent with the observation that more than 70% of incorporated entrepreneurs have a tertiary education degree, compared to less than half of the employees and self-employed. Less than 6% of the current incorporated entrepreneurs are female.

A correlation matrix appears in Table 2, based on the full sample. Ambiguity aversion is (negatively) correlated with ambiguity sensitivity. The finding that the two facets are correlated does not imply that they have the same influence on entrepreneurship, and our econometric analysis below reveals differential effects. As the two facets are not orthogonal, we simultaneously include them in our regressions to identify their unique effects.[11] We further find that ambiguity aversion is uncorrelated with risk aversion but negatively correlated with optimism, cognitive skill, and tertiary education. Younger individuals and females tend to be more ambiguity averse. The significant correlations between the ambiguity attitudes and the control variables suggest that controlling for these additional factors is potentially important ex ante, although a robustness check shows that the results concerning the ambiguity attitudes are not sensitive to dropping the controls.

[ Insert Table 2 about here. ]

Are attitudes toward ambiguity stable traits, or do they change, for example, due to entrepreneurial experience or exposure to an entrepreneurial environment? Due to the panel structure of the LISS, we can count the number of years a respondent has spent in the current occupation, unless the respondent had already been in this occupation when first entering the panel. Thus, for the subsamples of currently self-employed individuals and incorporated entrepreneurs who entered into this state while already participating in the LISS panel, we can estimate the correlation between the duration in this occupation and their ambiguity attitudes (Table B1 in Online Appendix B). The correlation coefficients are small and not statistically significantly different from zero. In other words, experience in self-employment or incorporated entrepreneurship does not appear to systematically alter individuals' attitudes toward ambiguity.

_______________

[11] The results are similar if we include ambiguity tolerance and ambiguity sensitivity separately in the regressions.

*5.2 Main model estimation results*

Table 3 shows the results from the main Probit regressions as average marginal effects.[12] Columns 1-4 use our main operationalization distinguishing between incorporated entrepreneurs and the self-employed. Columns 1 and 2 are based on the sample of working-age individuals, using their current employment status, whereas Columns 3 and 4 are based on the extended sample using both the current and previous employment status.

[ Insert Table 3 about here. ]

We find that ambiguity aversion is significantly and negatively associated with being an incorporated entrepreneur and that ambiguity sensitivity is significantly and positively associated with being self-employed, confirming our expectations.[13] These significant associations of the two components of ambiguity attitudes with the two types of entrepreneurs emerge while keeping other factors constant, in particular risk aversion, optimism, and cognitive skill. Contrary to our expectation, we do not find ambiguity sensitivity to be negatively associated with incorporated entrepreneurship. This suggests that, despite being more willing to bear ambiguity overall, incorporated entrepreneurs do not differ from employees in how much they adjust their actions in response to changes in their subjective probability assessments. Furthermore, risk aversion is negatively associated with both self-employment and incorporated entrepreneurship. These relationships are robustly found in both the sample of the working-age population as well as in the extended sample.

---

[12] Table B2 in Online Appendix B provides the corresponding Probit coefficients.
[13] We also test whether the coefficients of the ambiguity attitudes are significantly different for the choice outcomes self-employment and incorporated entrepreneur based on the multinomial logit model described in Section 5.3. The coefficients on ambiguity aversion are significantly different ($p$=0.013), but the coefficients on ambiguity sensitivity are not ($p$=0.250).

The marginal effect sizes on a given dependent variable are directly comparable across the first six independent variables because these are standardized. In the sample of working-age individuals, increasing ambiguity aversion by one standard deviation decreases the probability of being an incorporated entrepreneur by one percentage point (Column 2). Given that the share of incorporated entrepreneurs in this sample is 1.7% (as indicated at the bottom of the table), this corresponds to a relative effect size of 59%. In the same sample, a one standard deviation increase in ambiguity sensitivity is associated with a 2.2 percentage points higher probability of being self-employed (Column 1), corresponding to 31% of the average self-employment rate of 7.0%.

These effect sizes are comparable to those of the best-known determinants of entrepreneurship. For example, in Column 2, the point estimate of the significant effect of a one-standard-deviation change in ambiguity aversion is only slightly smaller than that of a one-standard-deviation change in cognitive skill in absolute terms. In the same column, the marginal effect of ambiguity aversion is also more than half the size (in absolute terms) of that of risk aversion, which is one of the most often discussed and studied preferences in entrepreneurship. This result is especially notable given that we are estimating partial effects while keeping the other variables constant, so the effect of ambiguity aversion is estimated in addition to and independent from the effect of risk aversion. In Column 2, the negative marginal effect of a one-standard-deviation increase in ambiguity aversion on the probability of being an incorporated entrepreneur is also as large as the positive effect of 10 years of aging, a proxy for experience, and a common predictor of entrepreneurship. Similarly, Azoulay et al. (2020) estimate that going from age 30 to 40 increases the probability of starting a highly successful U.S. business (in the top 1% of the growth distribution) from approximately 2% to 3% or one percentage point. Our data pins down the one-standard-deviation increase in ambiguity aversion to have approximately the same effect size on incorporated entrepreneurship in absolute terms.

Columns 5-6 use strictly positive growth in the number of employees supervised by an entrepreneur as an alternative measure of growth entrepreneurship. The share of growth entrepreneurs in the sample using this definition is 1.3%, somewhat smaller than the share of incorporated entrepreneurs with 1.7% (see the bottom row of the table). The estimated marginal effects of ambiguity aversion and ambiguity sensitivity are similar to Columns 1-2, both in terms of the point estimates and statistical significance. This indicates robustness of the results with respect to the operationalization of growth entrepreneurs. In Columns 7-8, we use employers (entrepreneurs supervising at least one worker) versus nonemployers (entrepreneurs not supervising anybody) as another distinction. We find a positive and significant association of ambiguity sensitivity with the probability of being a nonemployer, which is consistent with the effect on self-employment found in Column 1, and the point estimate is also similar. The point estimate of the effect of ambiguity aversion on the probability of being an employer is negative, as in Column 2 using incorporated entrepreneurship, but not statistically significant. As argued, employer status is not our preferred measure of intentions to grow because it partially reflects the minimum efficient size of a business, so measurement error may attenuate the association with growth entrepreneurship in Column 8. Indeed, the share of employer entrepreneurs in the sample is 2.8%, which is larger than the share of incorporated entrepreneurs, so it is possible that the group of employers includes a meaningful number of entrepreneurs who are not growth oriented.

To compare our findings to results from prior literature, it is insightful to build up our empirical model step by step. We start by using the traditional measure of ambiguity aversion based on the Ellsberg urns experiment. Such an experiment was included in the LISS survey in 2010 (Dimmock et al., 2016). Columns 1 and 2 of Table B3 in Online Appendix B show that there is no significant association of this measure of ambiguity aversion with neither self-employment nor incorporated entrepreneurship in 2010. This agrees with results reported in Holm et al. (2013) and Koudstaal et al. (2016). However, when we use our preferred measure of ambiguity aversion based on the belief-hedging method applied to the stock market instead, we find a negative and significant association with incorporated entrepreneurship, and the point estimate is similar to our main estimate of this effect in Column 2 of Table 3. The null finding when using Ellsberg urns may be explained by attenuation bias toward zero due to noisy measurement in the artificial setting of the Ellsberg urns experiment, whereas our measure in relation to the stock market, a natural and important source of uncertainty, seems to reduce noise in the measurement and thereby the attenuation bias. Next, we add ambiguity sensitivity to the model (Columns 5-6). This additionally reveals the positive and significant association between ambiguity sensitivity and self-employment, a novel finding of our analysis. The significantly negative association between ambiguity aversion and incorporated entrepreneurship remains stable. To finalize the model, it only remains to add the control variables, which leads to the main model shown in Columns 1 and 2 in Table 3. The inclusion of the control variables does not change the estimated effects much, indicating robustness.

*5.3 Comparison to hired managers and further robustness checks*

As we have found significant associations of ambiguity aversion with incorporated entrepreneurship and employment growth as indicators of growth entrepreneurship, an interesting further question is whether there is a similar association of ambiguity aversion with managers hired by a company they do not own. Incorporated entrepreneurs and hired managers overlap with respect to many tasks they typically perform, including supervising others and planning strategically. However, there are also important differences, for example, entrepreneurs bear more risk personally, and they operate in a more uncertain environment with less data available in comparison to a manager in a larger organization with a longer track record (Koudstaal et al., 2019). To compare hired managers to incorporated entrepreneurs, we estimate a multinomial logit model with four discrete choice categories: hired manager, self-employed, incorporated entrepreneur, and non-managerial employee; the latter is the omitted base category. Hired managers are identified in the LISS data as employees who supervise at least one worker; employees who do not supervise anybody are defined as non-managerial employees. Table B4 in Online Appendix B shows the results. The coefficients can be approximately compared to the Probit coefficients in Table B2 (Columns 1-2) after multiplying them by 0.6 (Amemiya, 1981). This comparison, as well as the significance levels, shows that the estimated effects of the ambiguity parameters on self-employment and incorporate entrepreneurship are similar to the results from our main binary choice estimations, indicating robustness. Consistent with Koudstaal et al. (2016), we find that hired managers exhibit significantly lower risk aversion than non-managerial employees. Interestingly, the coefficients of the ambiguity attitudes are not significant for the choice to be a hired manager in comparison to being a non-managerial employee. This indicates that ambiguity attitudes are unique determinants of entrepreneurship while not predicting the probability of being a hired manager.

We conduct three more sets of robustness checks to assess the sensitivity of our results with respect to the definitions of the treatment and control groups, the sample, and the outcome variables. For these tests, we return to the binary Probit models. First, our distinction between different types of entrepreneurs is related to the literature on opportunity-oriented versus necessity-oriented entrepreneurship (e.g., Margolis, 2014; Dencker et al., 2021, O'Donnell et al., 2024). To assess whether our results are driven by necessity-based entrepreneurship, we redefine our outcome variables to exclude necessity entrepreneurs in this robustness check. We operationalize necessity entrepreneurs as those who were last observed in unemployment in the LISS panel before they became entrepreneurs (Block and Wagner, 2010; Fairlie and Fossen, 2020). The estimated marginal effects shown in Table B5 (Columns 1-2) remain very similar to our main results in Table 3 (Columns 1-2), suggesting that our main results are not due to a dominating influence of necessity entrepreneurs. Second, we exclude incorporated entrepreneurs from the sample as part of the comparison group when estimating the probability of self-employment and vice versa. The results in Columns 3-4 of Table B5 indicate robustness. Third, the results are also stable when we exclude on-call employees and temp staffers from the sample (Columns 5-6 of Table B5).

## 6. Discussion

Entrepreneurs deal with high levels of ambiguity where it is not possible to precisely estimate the probabilities of future outcomes. In such situations, both the entrepreneurs' subjective beliefs and attitudes toward ambiguity are key. The role of subjective beliefs—the perception of the likelihood that an event will occur—has been highlighted before (Koellinger et al., 2007; Felin and Zenger, 2009; Chen et al., 2018; Zellweger and Zenger, 2023; Camuffo et al., 2024; Gans, 2024). Just as crucial, attitude toward ambiguity is a fundamental element of several theories of entrepreneurial decision-making (McMullen and Shepherd, 2006; McKelvie et al., 2011; Foss and Klein, 2012).

We advance understanding of entrepreneurial decision-making by proposing that two dimensions of ambiguity attitudes affect entrepreneurial action. Given beliefs, action depends not only on the willingness to bear ambiguity but also on how finely individuals discriminate among perceived chances of success. This second dimension of ambiguity attitudes—ambiguity sensitivity—is conceptually different from tolerance for ambiguity. Using incentivized measures from a large population survey, we show the importance of separating these two dimensions of ambiguity attitudes as different types of entrepreneurs exhibit systematically different patterns across the two components.

*6.1 Contributions to the literature*

This study makes five contributions to the literature. First, we advance the literature on ambiguity attitudes in entrepreneurship. While entrepreneurship scholars have long emphasized ambiguity attitudes in entrepreneurial action (Frenkel-Brunswick, 1951; Teoh and Foo, 1997; Mueller and Thomas, 2001; McMullen and Shepherd, 2006), empirical evidence has been inconsistent, possibly due to conceptual and methodological challenges. Conceptually, scholars have highlighted the lack of clear separation between risk and ambiguity attitudes (Furnham and Marks, 2013). Methodologically, existing studies that separate ambiguity attitudes from risk attitudes using behavioral data (e.g., Holm et al., 2013; Koudstaal et al., 2016) have primarily focused on tolerance for ambiguity and relied on Ellsberg urn-type experiments, which, while theoretically elegant, may have limited predictive power for real-world decision-making contexts. We address these challenges by distinguishing two dimensions of ambiguity attitudes using incentivized decisions involving financial assets with real-world relevance. We clarify that tolerance for ambiguity captures the willingness to pay for having ambiguity reduced, as compared to having risk reduced, thus establishing a clear conceptual distinction from risk tolerance. We further distinguish ambiguity sensitivity from tolerance for ambiguity, where ambiguity sensitivity describes how individuals discriminate between different levels of perceived chances of success, a concept that has until now remained unexplored in entrepreneurship research. Ambiguity sensitivity is particularly important for entrepreneurial decision-making as it determines how entrepreneurial beliefs about opportunities translate into action, offering a new perspective on entrepreneurial judgment and decision-making.

Second, we contribute to the literature on entrepreneurial heterogeneity (Minniti and Lévesque, 2010; Kuechle, 2011; Levine and Rubinstein, 2017) by theorizing that entrepreneurs with different growth intentions differ systematically in how they respond to ambiguity. We measure growth intentions through business incorporation, although a measure based on actual growth level yields consistent conclusions. Incorporated entrepreneurs targeting fast-growth opportunities need to enlist much more external resources (e.g., venture capital) than the self-employed (who typically rely on personal savings, family, or bank loans). These differing external environments sort individuals into two different classes. Incorporated entrepreneurs need to demonstrate "persistence" (Yan et al., 2023), use coherent storytelling (Garud et al., 2014), and be persuasive (Chen et al., 2009; Hu and Ma, 2025; Stefanidis et al., 2025) to secure external resources. We theorize that they are therefore, inherently, more tolerant to ambiguity but are less ambiguity sensitive; they resist changing their story and actions when perceived probabilities change. The self-employed, on the other hand, are expected to adjust their actions more strongly when market signals change as they rely on observable information in relatively predictable environments to survive financially. Our empirical findings confirm that incorporated entrepreneurs show greater tolerance for ambiguity than employees, whereas the self-employed exhibit higher ambiguity sensitivity; however, incorporated entrepreneurs do not differ from employees concerning ambiguity sensitivity. While incorporated entrepreneurs are known to differ from the self-employed in their growth and entrepreneurial and innovative intentions (Hurst and Pugsley, 2011; Åstebro and Tåg, 2017; Levine and Rubinstein, 2017; Estrin et al., 2024), we show that they also differ fundamentally in how their perceptions of opportunities translate into action.

Third, our findings also speak to the literature on entrepreneurial judgment and learning. For instance, a growing literature emphasizes how entrepreneurs can update their beliefs through experimentation (Chen et al., 2018, 2024; Ehrig and Schmidt, 2022; Zellweger and Zenger, 2023; Camuffo et al., 2024; Cao et al., 2024). This literature implicitly assumes that entrepreneurs will act differently only based on their updated probability assessments. However, we highlight that ambiguity sensitivity is a critical dimension that determines to what extent belief updating translates into entrepreneurial action. If some entrepreneurs do not substantially alter their actions in response to changes in their beliefs, training these entrepreneurs to better update their beliefs may have limited effects. Given that the self-employed exhibit higher ambiguity sensitivity, implying that they react more strongly to new information, institutions in the entrepreneurial ecosystem may be able to support this group effectively by providing market information. These insights emphasize the importance of considering attitudes toward ambiguity in theories of entrepreneurial judgment and learning.

Fourth, at the conceptual level, we propose an alternative explanation for why some entrepreneurs appear to commit to a decision without revising it, even when presented with new information. This behavior has been described in various ways, such as "persistence" (Yan et al., 2023)[14], "optimism" (Tenney et al., 2015) and "overconfidence" (Moore and Schatz, 2017). We introduce a framework that explains such decision patterns through differences in ambiguity sensitivity—the reluctance or proclivity to change one's course of action as a function of updated beliefs. By doing so, our framework helps explain why some entrepreneurs adapt quickly to new evidence while others stay committed to their initial course of action, speaking to parts of the literature on persistence, optimism, and overconfidence.

---

[14] For example, Shane et al. (2003, p. 260) define "people who are willing to proceed despite [unfavorable] odds" as persistent. For a review of the many ways that the concept has been presented, see Yan et al. (2023).

Fifth, our findings also highlight a potential misalignment between entrepreneurs and their investors. While growth-oriented entrepreneurs exhibit a relatively large tolerance for ambiguity, many investors do not (Easley and O'Hara, 2009; Anantanasuwong et al., 2024). This fundamental divergence in ambiguity attitudes can create friction in the investor-entrepreneur relationship. For instance, investors may interpret ambiguity insensitive entrepreneurial decisions as optimistic, overconfident or having poor judgment, while the entrepreneur may be driven by a feeling that they need to show persuasiveness. While this is beyond the scope of this study, such misalignment of incentives could lead to disagreements, stalemates, and, for example, erroneous investments or project terminations by the investor (e.g., Van den Steen, 2004). In articles by Van den Steen (2004; 2010; 2011) such disagreements and incentive misalignments are explained by rational agents having differing priors, i.e. different beliefs. In this article we raise the possibility that different actors may also have different ambiguity attitudes, which may cause misaligned actions even if beliefs are identical. Further theoretical and empirical analysis is necessary to reveal the implications of such differing ambiguity attitudes. Investors may need to adapt governance mechanisms to better align entrepreneurial actions with their investment preferences.

### 6.2 Limitations and outlook

Our analysis relies on high-quality, incentive-compatible measures of ambiguity attitudes in a general population survey (LISS). However, our study has some limitations that suggest directions for future research. First, our methodology assumes that ambiguity aversion and ambiguity sensitivity are stable personality traits. While it is possible that the two components of ambiguity attitudes may vary over time, current evidence suggests that ambiguity attitudes are relatively stable (Duersch et al., 2017; Von Gaudecker et al., 2022). Consistent with this, we find no significant correlation between the time spent in self-employment or as an incorporated entrepreneur with the ambiguity attitudes of the entrepreneurs, suggesting that experience from entrepreneurship does not change ambiguity attitudes. In any case, due to the cross-sectional nature of our ambiguity data, we do not claim causality. It remains unclear whether individuals self-select into environments that correspond with their pre-existing ambiguity attitudes, whether entrepreneurial experience shapes ambiguity attitudes, or whether both processes operate concurrently. Subsequent studies employing longitudinal designs or experimental methods could further elucidate whether ambiguity aversion and sensitivity are stable traits that predispose individuals to particular entrepreneurial paths or if they are, at least in part, outcomes of entrepreneurial experiences.

Second, our measures of ambiguity attitudes are based on incentivized decisions in financial markets, yet entrepreneurs face uncertainty across multiple domains. It may be possible that individuals exhibit different attitudes towards ambiguity in different decision situations. For example, Abdellaoui et al. (2011) find different levels of tolerance for ambiguity across different domains of uncertainty but no significant differences in ambiguity sensitivity. In contrast, some studies suggest relatively stable ambiguity attitudes across different sources of uncertainty (e.g., Von Gaudecker et al., 2022; Anantanasuwong et al., 2024). Future research using measures of ambiguity attitudes elicited over different outcomes (e.g., new product success or market entry) could test whether the patterns we observe generalize to other decision environments entrepreneurs navigate.

Third, our research identifies differences in attitudes toward ambiguity between two types of entrepreneurs and employees but does not examine other entrepreneurial outcomes. Future research could explore how ambiguity aversion and ambiguity sensitivity affect the different stages of entrepreneurial decision-making, from opportunity recognition to exit decisions and, ultimately, entrepreneurial performance. For example, an interesting area for theoretical exploration is to what extent differences in ambiguity attitudes between investors and entrepreneurs affect project termination decisions. In addition, our finding that self-employed individuals exhibit elevated levels of ambiguity sensitivity suggests that they might be better equipped than incorporated entrepreneurs to learn from experience and new information and adjust their decisions accordingly.

Fourth, this study focuses on situations of ambiguity, when entrepreneurs can form subjective probabilities. Research on situations of ignorance, when entrepreneurs cannot even conceive the possible future states and consequently cannot assign subjective probabilities, and on how entrepreneurs transition from such situations to situations of ambiguity, remains critically important.

**Tables**

**Table 1: Descriptive statistics by employment category**

| | Currently | | | | | |
| | Employee | | Self-employed | | Incorp. entrepreneur | |
| Variable | Mean | SD | Mean | SD | Mean | SD |
|---|---|---|---|---|---|---|
| Ambiguity aversion | 0.010 | 0.961 | 0.030 | 1.003 | -0.630 | 0.626 |
| Ambiguity sensitivity | 0.012 | 0.992 | 0.330 | 1.024 | 0.472 | 0.929 |
| Risk aversion | -0.028 | 0.975 | -0.352 | 0.948 | -0.931 | 0.685 |
| Optimism | 0.007 | 1.028 | -0.061 | 1.112 | 0.380 | 0.962 |
| Cognitive skill | 0.082 | 0.975 | 0.203 | 1.027 | 0.782 | 0.403 |
| Upper secondary education | 0.391 | | 0.378 | | 0.167 | |
| Tertiary education | 0.454 | | 0.486 | | 0.778 | |
| Age | 46.00 | 11.85 | 49.81 | 10.14 | 51.39 | 10.46 |
| Female | 0.507 | | 0.527 | | 0.056 | |
| Married | 0.461 | | 0.581 | | 0.611 | |
| No. children in household | 0.786 | 1.061 | 0.973 | 1.314 | 0.722 | 1.127 |
| N | 970 | | 74 | | 18 | |
| | Ever observed as | | | | | |
| | Employee | | Self-employed | | Incorp. entrepreneur | |
| Variable | Mean | SD | Mean | SD | Mean | SD |
| Ambiguity aversion | 0.005 | 0.999 | 0.045 | 1.046 | -0.400 | 0.701 |
| Ambiguity sensitivity | -0.022 | 0.999 | 0.127 | 1.010 | 0.221 | 0.918 |
| Risk aversion | 0.041 | 1.005 | -0.189 | 0.938 | -0.633 | 0.793 |
| Optimism | -0.023 | 0.997 | 0.121 | 1.043 | 0.291 | 0.791 |
| Cognitive skill | -0.032 | 1.006 | 0.166 | 0.970 | 0.405 | 0.712 |
| Upper secondary education | 0.353 | | 0.317 | | 0.200 | |
| Tertiary education | 0.394 | | 0.467 | | 0.700 | |
| Age | 55.10 | 15.46 | 57.70 | 14.21 | 59.30 | 13.97 |
| Female | 0.486 | | 0.508 | | 0.200 | |
| Married | 0.507 | | 0.513 | | 0.525 | |
| No. children in household | 0.515 | 0.926 | 0.608 | 1.095 | 0.450 | 0.904 |
| N | 1543 | | 199 | | 40 | |

*Notes*: The upper panel shows the sample of working-age individuals (21-64 years of age), the lower panel the extended sample (21 years of age or older). The "self-employed" are self-employed, freelancers, or independent professionals, "incorporated entrepreneurs" are directors of limited liability or private limited companies or majority shareholder directors. The variables in the first six rows of each panel are standardized based on the extended sample used in the second panel with 1782 observations. SD is standard deviation (not shown for binary variables). *Source*: Own calculations based on the LISS.

**Table 2: Correlation matrix**

|  |  | 1 | 2 | 3 | 4 | 5 | 6 | 7 | 8 | 9 | 10 | 11 |
|---|---|---|---|---|---|---|---|---|---|---|---|---|
| 1 | Ambiguity aversion | 1.000 | | | | | | | | | | |
| 2 | Ambig. sensitivity | -0.316 | 1.000 | | | | | | | | | |
|  |  | 0.000 | | | | | | | | | | |
| 3 | Risk aversion | 0.021 | -0.047 | 1.000 | | | | | | | | |
|  |  | 0.387 | 0.048 | | | | | | | | | |
| 4 | Optimism | -0.071 | 0.099 | -0.117 | 1.000 | | | | | | | |
|  |  | 0.003 | 0.000 | 0.000 | | | | | | | | |
| 5 | Cognitive skill | -0.182 | 0.305 | -0.042 | 0.234 | 1.000 | | | | | | |
|  |  | 0.000 | 0.000 | 0.076 | 0.000 | | | | | | | |
| 6 | Upper sec. edu. | 0.014 | -0.084 | 0.017 | -0.066 | -0.047 | 1.000 | | | | | |
|  |  | 0.564 | 0.000 | 0.472 | 0.005 | 0.048 | | | | | | |
| 7 | Tertiary educ. | -0.102 | 0.217 | -0.087 | 0.196 | 0.358 | -0.604 | 1.000 | | | | |
|  |  | 0.000 | 0.000 | 0.000 | 0.000 | 0.000 | 0.000 | | | | | |
| 8 | Age | -0.021 | -0.044 | 0.124 | 0.056 | -0.105 | -0.096 | -0.165 | 1.000 | | | |
|  |  | 0.367 | 0.063 | 0.000 | 0.018 | 0.000 | 0.000 | 0.000 | | | | |
| 9 | Female | 0.087 | -0.156 | 0.137 | -0.043 | -0.229 | 0.034 | -0.095 | -0.133 | 1.000 | | |
|  |  | 0.000 | 0.000 | 0.000 | 0.068 | 0.000 | 0.147 | 0.000 | 0.000 | | | |
| 10 | Married | 0.002 | 0.050 | 0.037 | 0.098 | 0.032 | 0.054 | -0.064 | 0.207 | -0.158 | 1.000 | |
|  |  | 0.941 | 0.034 | 0.116 | 0.000 | 0.175 | 0.024 | 0.007 | 0.000 | 0.000 | | |
| 11 | No. of children | -0.020 | 0.017 | -0.026 | 0.025 | 0.036 | 0.098 | 0.023 | -0.269 | 0.038 | 0.226 | 1.000 |
|  |  | 0.400 | 0.480 | 0.269 | 0.299 | 0.126 | 0.000 | 0.336 | 0.000 | 0.111 | 0.000 | |

*Notes*: The table shows pairwise correlation coefficients and below *p*-values for tests of whether a correlation coefficient differs from zero. Based on the sample of 1,782 observations with age above 20. *Source*: Own calculations based on the LISS.

**Table 3: Probit estimations – average marginal effects**

| Dependent variable: | (1) Currently | (2) | (3) Ever observed as | (4) | (5) Currently entrepreneur with | (6) | (7) Currently | (8) |
|---|---|---|---|---|---|---|---|---|
| | Self-employed | Incorp. entrepreneur | Self-employed | Incorp. entrepreneur | No employment growth | Employment growth | Non-employer | Employer |
| Ambiguity aversion | 0.012 | -0.010** | 0.016* | -0.007** | 0.008 | -0.011** | 0.008 | -0.004 |
| | (0.009) | (0.004) | (0.008) | (0.003) | (0.009) | (0.004) | (0.009) | (0.006) |
| Ambiguity sensitivity | 0.022*** | 0.000 | 0.015* | -0.001 | 0.028*** | 0.000 | 0.019** | 0.002 |
| | (0.009) | (0.004) | (0.008) | (0.003) | (0.009) | (0.003) | (0.008) | (0.005) |
| Risk aversion | -0.026*** | -0.017*** | -0.026*** | -0.014*** | -0.030*** | -0.013** | -0.023*** | -0.019*** |
| | (0.008) | (0.005) | (0.008) | (0.004) | (0.009) | (0.005) | (0.008) | (0.005) |
| Optimism | -0.015* | -0.001 | 0.003 | 0.002 | -0.014* | -0.001 | -0.010 | -0.006 |
| | (0.008) | (0.004) | (0.008) | (0.003) | (0.008) | (0.004) | (0.008) | (0.005) |
| Cognitive skill | 0.007 | 0.014** | 0.020** | 0.004 | 0.002 | 0.008 | 0.003 | 0.013* |
| | (0.010) | (0.006) | (0.010) | (0.004) | (0.010) | (0.006) | (0.009) | (0.007) |
| Upper sec. educ. | 0.016 | -0.004 | 0.007 | 0.004 | 0.008 | 0.004 | 0.025 | -0.006 |
| | (0.025) | (0.013) | (0.022) | (0.014) | (0.025) | (0.014) | (0.025) | (0.014) |
| Tertiary education | 0.024 | 0.012 | 0.021 | 0.023* | 0.027 | 0.010 | 0.036 | 0.006 |
| | (0.025) | (0.012) | (0.023) | (0.012) | (0.026) | (0.012) | (0.025) | (0.014) |
| Age | 0.002*** | 0.001** | 0.002*** | 0.001** | 0.003*** | 0.000 | 0.002*** | 0.001** |
| | (0.001) | (0.000) | (0.001) | (0.000) | (0.001) | (0.000) | (0.001) | (0.000) |
| Female | 0.032* | -0.015*** | 0.039** | -0.014** | 0.017 | -0.004 | 0.017 | 0.001 |
| | (0.017) | (0.006) | (0.016) | (0.006) | (0.017) | (0.008) | (0.015) | (0.010) |
| Married | 0.017 | 0.010 | -0.012 | -0.002 | 0.028* | 0.002 | 0.030** | -0.003 |
| | (0.016) | (0.008) | (0.016) | (0.007) | (0.016) | (0.007) | (0.015) | (0.010) |
| No. of children | 0.007 | -0.002 | 0.018** | -0.001 | 0.005 | 0.000 | -0.006 | 0.009** |
| | (0.007) | (0.003) | (0.008) | (0.004) | (0.007) | (0.003) | (0.007) | (0.004) |
| N | 1062 | 1062 | 1782 | 1782 | 1000 | 1000 | 1062 | 1062 |
| Log likelihood | -251.575 | -65.404 | -604.178 | -167.813 | -230.939 | -56.246 | -220.120 | -122.224 |
| Mean dep. variable | 0.070 | 0.017 | 0.112 | 0.022 | 0.070 | 0.013 | 0.058 | 0.028 |

*Notes*: Average marginal effects obtained from Probit estimations. Age above 20 in all columns and also below 65 in all columns except (3) and (4). The first six independent variables are standardized. Robust standard errors are shown in parentheses. ***/**/* indicate statistical significance at the 1%/5%/10% level. *Source*: Own calculations based on the LISS.

# Appendix A: Method to estimate ambiguity attitudes in a natural context

## A.1 The belief-hedging method: core intuition

We use the belief-hedging method to estimate ambiguity aversion and ambiguity sensitivity in the natural context of stock markets. At its core, the belief-hedging method leverages bets on complementary events—pairs of events that cover all possible outcomes of a scenario, such as a stock price either rising or falling—to measure ambiguity attitudes. It exploits a fundamental principle: while we may not know an individual's beliefs about specific events, we know that the sum of the subjective probabilities of two complementary events equals 100%. For example, an investor might believe there is a 60% likelihood of a stock price increase and a 40% likelihood of a decrease. By examining bets placed across a spectrum of complementary events, the method assesses not just the willingness to bet on individual events, which is influenced by both beliefs and attitudes, but also the aggregate willingness, which reflects solely the decision-maker's attitudes toward ambiguity.

We use events related to the returns from investing in the Amsterdam Exchange Index (AEX), the most widely known stock market index in the Netherlands; more specifically, we use the returns from investing 1000 euros in the AEX six months post-investment. The method uses a partition of three mutually exclusive and exhaustive events: the investment's value dropping below 950 euros, ranging between 950 and 1100 euros, and exceeding 1100 euros. We denote these events as $E_l = (-\infty, 950)$, $E_m = [950, 1100]$, and $E_h = (1100, \infty)$, where l, m, and h stand for low, medium, and high returns, respectively. The method also uses the complementary events to these three 'singular' events. These events are the union (composite) of the two other events. For instance, the complementary event to $E_h$ is $E_{lm} = E_l \cup E_m = (-\infty, 1100]$, i.e., the union of $E_l$ and $E_m$, which means that the value of the investment is below 1100 euros.

In each wave, the survey measures participants' matching probabilities for each of six events: the three singular and the three composite events. The matching probability of event $E$ is the objective probability $q$ that makes the participant indifferent between a bet awarding 20 euros with probability $q$ and a bet yielding 20 euros if event $E$ occurs. For instance, if a participant is indifferent between a bet that pays 20 euros with an 80% chance and a bet that pays 20 euros if the investment return exceeds 1,100 euros (event $E_h$), then the matching probability of the event $E_h$ is 80%.

This method allows for the direct measurement of 'local' ambiguity aversion without necessitating the quantification of the utility function or risk aversion levels (Dimmock et al., 2016). For instance, if an individual attributes a subjective probability of 80% to event $E_h$ but assigns a matching probability of 60% for the same event, this suggests local ambiguity aversion. The decision maker is as comfortable betting on an ambiguous event with an 80% subjective probability as she is on a risky event with a 60% chance of winning, indicating ambiguity aversion. Risk attitude cannot capture this preference pattern, as risk attitude treats subjective probabilities the same as objective probabilities.

Consider another scenario where the matching probability for the complementary event $E_{lm}$, i.e., the return on the investment is at or below 1100 euros, is 30%. This behavior indicates local ambiguity tolerance, as the matching probability of 30% is higher than the subjective probability of 20% for this event, where 20% is the complement of the 80% subjective probability assigned to $E_h$.

While these examples assume that we know the subjective probabilities assigned to events, the belief-hedging method enables assessing an overall ambiguity tolerance level without requiring direct measurement of subjective beliefs. It does so by computing the sum of the matching probabilities for two complementary events. Given that the sum of subjective probabilities for two complementary events (for instance, $E_h$ and $E_{lm}$) equals 100%, a sum of the matching probabilities for these events below 100%—for instance, reaching only 60% + 30% = 90% in the previous example—indicates a general ambiguity aversion. Conversely, if the sum of the two matching probabilities exceeds 100%, it indicates a tolerance for ambiguity.[16]

The second dimension of ambiguity attitudes, ambiguity sensitivity, reflects how individuals differentiate between varying levels of perceived chances of success. Essentially, low ambiguity sensitivity implies a similar response to ambiguous situations even if these have different subjective probabilities. The belief-hedging method evaluates ambiguity sensitivity by comparing the matching probabilities of singular events against those of their complementary counterparts. Since the sum of subjective beliefs for the three singular events totals 100%, and that for the three complementary events totals 200% due to the overlap of the composite events, an individual with neutral ambiguity sensitivity should have a cumulative matching probability for complementary events that is double that of the singular events. In other words, such an individual's actions reflect that, on average, complementary events are twice as likely to occur as singular events. Instead, an individual exhibiting low ambiguity sensitivity tends to wager relatively more on lower-probability events and relatively less on higher-probability ones. Consequently, for such individuals, the sum of matching probabilities for complementary events will be less than twice the sum of matching probabilities for singular events.

*A.2 Measurement of the ambiguity attitudes*

The ambiguity module in the LISS survey implements the belief-hedging method to elicit ambiguity attitudes (Baillon et al., 2018; Baillon et al., 2021). As explained above, respondents make a series of monetarily incentivized choices related to hypothetical investments in the AEX. In each task, individuals are asked to choose between receiving 20 euros if an ambiguous event occurs (such as: an investment of 1,000 euros in the AEX will be worth more than 1,100 euros in six months) and a chance of winning 20 euros in a lottery with some known probability. For each ambiguous event, the computerized questionnaire navigates the respondent through a decision tree offering different winning probabilities in the lottery, subsequently narrowing down the bounds of the individual's matching probability for the event. The matching probability makes the individual indifferent between receiving 20 euro with that probability in the lottery and receiving 20 euro if the ambiguous event occurs.

---

[16] Thus, the ambiguity model assumes additivity of subjective probabilities but allows decision weights and matching probabilities to be non-additive. This allows to clearly separate ambiguity attitudes from beliefs (Machina and Schmeidler, 1992; Wakker, 2004; Abdellaoui et al., 2024).

Paying out the result from each choice to the subjects would be prohibitively expensive. Following standard practice in incentivized experiments, a random question was selected for payout. For example, if a subject chose a lottery in this question, the lottery was played out; if the subject chose to bet on the AEX, the evolution of the AEX was checked after six months to determine the payout. Expected incentive payments per respondent were 13.50 euros, implying a substantial hourly wage rate of 51 euros given the median response time. Subjects started a random number generator to select a question to be paid out before they made any decisions to preserve incentive compatibility (Bardsley, 2000; Johnson et al., 2021; Von Gaudecker et al., 2022).

Based on the choice data, we follow Von Gaudecker et al. (2022) and use a structural stochastic choice model to estimate three parameters for each individual in the sample: ambiguity aversion, ambiguity sensitivity, and the magnitude of decision errors. The first two are the main explanatory variables of interest in our regressions. In the following, we outline the estimation strategy for the three parameters; Von Gaudecker et al. (2022) provide further details.

The first two parameters are defined as:

Ambiguity aversion $= E[Prob_{subj}(event) - W(event)]$

Ambiguity sensitivity $= \frac{Cov[W(event), Prob_{subj}(event)]}{Var[Prob_{subj}(event)]}$

where $Prob_{subj}(event)$ is the subjective probability of the ambiguous event, and $W(event)$ is the decision weight (Abdellaoui et al., 2011). If the subjective probabilities are equal to the decision weights, individuals maximize their subjective expected utility, which means that they are ambiguity neutral. However, if the subjective probabilities are larger than the decision weights, then individuals are ambiguity averse; if the subjective probabilities are smaller, individuals are ambiguity seeking. Ambiguity sensitivity captures how much decision weights move with changes in subjective probabilities. If decision weights are perfectly correlated with subjective probabilities, ambiguity sensitivity is one. The lower the correlation, the less sensitive the decision maker is to changes in subjective probabilities.

Von Gaudecker et al. (2022) further assume that the decision weights, $W(event)$, are non-extreme outcome additive (Chateauneuf et al., 2007). This means that decision weights range between 0 and 1 and are linear functions of subjective probabilities within this interval. This functional form has the advantages that it is tractable and shows good empirical performance (Li et al., 2018).

The matching probabilities measured from the choices, $m(event)$, may differ from the theoretical decision weights due to normally distributed decision errors, $e$:

Decision error: $m(event) = W(event) + e$ with $e \sim N(0, \sigma^2)$

This stochastic component of the choice model accommodates set-monotonicity violations, i.e., choices that reveal higher matching probabilities for an event that is a strict subset of another event (Von Gaudecker et al., 2022). This occurs, for example, when an individual is indifferent between receiving 20 euros with 50% chance and receiving 20 euros if an investment of 1000 euros in the AEX will be worth more than 950 euros in six months, and the same individual is also indifferent between receiving 20 euros with 70% chance (which is clearly better) and receiving 20 euros if the investment will be worth more than 1100 euros (which is clearly worse). Such inconsistent choices are frequently observed empirically but cannot be rationalized by purely deterministic theories of choice under uncertainty. The decision error also captures individual choice behavior that is erratic across survey waves. The approach of modeling decision error as a stochastic model component is superior to alternative approaches such as simply dropping individuals with invalid choices or keeping all observations regardless of whether the estimated parameters make sense. In a robustness check (available from the authors on request), we included the estimated standard deviation of the decision error for each individual as an additional control variable in our main regressions, and the results remained almost identical.

For each ambiguous event, completion of all choice tasks yields a lower bound, *lb(event)*, and an upper bound, *ub(event)*, of the individual's matching probability. These observed bounds of the matching probabilities are then used to estimate the parameters of the structural stochastic choice model using the maximum likelihood method for each respondent. Each individual in the sample participated in 2-6 survey waves eliciting ambiguity attitudes; we pool all these waves for our estimation. The likelihood function for each individual is written as

$$L(parameters) = \prod_{events} Prob[lb(event) \leq m(event) \leq ub(event)]$$

and maximized subject to model constraints, which include that the structural parameters do not vary across survey waves.

Von Gaudecker et al. (2022) show that the two ambiguity parameters (ambiguity aversion and ambiguity sensitivity) are very heterogeneous in the population. This motivates our approach of relating individual ambiguity attitudes to differences in entrepreneurial activity. Von Gaudecker et al. (2022) also analyze the panel dimension in the choice data and document that the parameters are stable over time. Further, they compare the ambiguity parameters elicited in the domain of the stock market, which we use here, to ambiguity parameters elicited in the domain of climate change, and they find that the parameters are largely stable across domains as well. The documented stability of the ambiguity parameters increases confidence that these parameters reflect meaningful personal traits.

# Appendix B: Supplemental tables

**Table B1: Duration of entrepreneurship and ambiguity attitudes**

|  | (1) Self-employed | (2) Incorp. entrepreneur |
|---|---|---|
| Ambiguity aversion | -0.079 | 0.093 |
| *p*-value | 0.516 | 0.786 |
| Ambiguity sensitivity | 0.013 | 0.040 |
| *p*-value | 0.915 | 0.908 |
| *N* | 70 | 11 |

*Notes*: Correlation coefficients between the ambiguity attitudes and the duration in years that individuals have spent in their currently ongoing spell of self-employment or as an incorporated entrepreneur. Below: *p*-values pertaining to tests whether the correlations are zero. *Source*: Own calculations based on the LISS.

**Table B2: Probit estimations – coefficients**

| Dependent variable: | (1) Currently | (2) | (3) Ever observed as | (4) | (5) Currently entrepreneur with | (6) | (7) Currently | (8) |
|---|---|---|---|---|---|---|---|---|
| | Self-employed | Incorp. entrep. | Self-employed | Incorp. entrep. | No employment growth | Employment growth | Non-employer | Employer |
| Ambiguity aversion | 0.098 | -0.309** | 0.087* | -0.149** | 0.066 | -0.377*** | 0.070 | -0.063 |
| | (0.074) | (0.121) | (0.045) | (0.062) | (0.076) | (0.139) | (0.078) | (0.095) |
| Ambiguity sensitivity | 0.179*** | 0.015 | 0.081* | -0.021 | 0.228*** | 0.001 | 0.170** | 0.042 |
| | (0.068) | (0.120) | (0.044) | (0.070) | (0.071) | (0.118) | (0.073) | (0.082) |
| Risk aversion | -0.210*** | -0.544*** | -0.141*** | -0.282*** | -0.240*** | -0.457*** | -0.210*** | -0.330*** |
| | (0.065) | (0.111) | (0.042) | (0.071) | (0.068) | (0.166) | (0.072) | (0.073) |
| Optimism | -0.120* | -0.032 | 0.015 | 0.035 | -0.118* | -0.050 | -0.090 | -0.099 |
| | (0.063) | (0.130) | (0.044) | (0.068) | (0.068) | (0.126) | (0.068) | (0.081) |
| Cognitive skill | 0.057 | 0.450*** | 0.107** | 0.084 | 0.015 | 0.268 | 0.023 | 0.225* |
| | (0.079) | (0.171) | (0.054) | (0.091) | (0.082) | (0.215) | (0.083) | (0.123) |
| Upper sec. educ. | 0.123 | -0.145 | 0.037 | 0.090 | 0.067 | 0.147 | 0.220 | -0.107 |
| | (0.189) | (0.467) | (0.118) | (0.266) | (0.197) | (0.430) | (0.207) | (0.262) |
| Tertiary education | 0.187 | 0.416 | 0.113 | 0.472* | 0.214 | 0.348 | 0.324 | 0.103 |
| | (0.193) | (0.455) | (0.120) | (0.246) | (0.203) | (0.439) | (0.214) | (0.246) |
| Age | 0.020*** | 0.024** | 0.012*** | 0.012** | 0.025*** | 0.003 | 0.020*** | 0.017** |
| | (0.006) | (0.011) | (0.003) | (0.006) | (0.006) | (0.010) | (0.006) | (0.008) |
| Female | 0.252* | -0.632** | 0.208** | -0.310** | 0.141 | -0.147 | 0.154 | 0.019 |
| | (0.129) | (0.319) | (0.087) | (0.156) | (0.132) | (0.306) | (0.131) | (0.175) |
| Married | 0.135 | 0.319 | -0.066 | -0.049 | 0.232* | 0.083 | 0.277** | -0.049 |
| | (0.129) | (0.230) | (0.087) | (0.154) | (0.131) | (0.258) | (0.133) | (0.177) |
| No. of children | 0.053 | -0.077 | 0.096** | -0.014 | 0.042 | 0.008 | -0.055 | 0.157** |
| | (0.058) | (0.106) | (0.045) | (0.080) | (0.060) | (0.093) | (0.068) | (0.064) |
| Constant | -2.879*** | -4.154*** | -2.090*** | -2.931*** | -3.130*** | -3.103*** | -3.004*** | -3.040*** |
| | (0.355) | (0.647) | (0.224) | (0.466) | (0.404) | (0.657) | (0.384) | (0.450) |
| N | 1062 | 1062 | 1782 | 1782 | 1000 | 1000 | 1062 | 1062 |
| Log likelihood | -251.575 | -65.404 | -604.178 | -167.813 | -230.939 | -56.246 | -220.120 | -122.224 |
| Mean dep. variable | 0.070 | 0.017 | 0.112 | 0.022 | 0.070 | 0.013 | 0.058 | 0.028 |

*Notes*: Probit coefficients. Age above 20 in all columns and also below 65 in all columns except (3) and (4). The first six independent variables are standardized. Robust standard errors in parentheses. ***/**/* indicate statistical significance at the 1%/5%/10% level. *Source*: Own calculations based on the LISS.

**Table B3: Different measures of ambiguity attitudes – average marginal effects**

| | (1) | (2) | (3) | (4) | (5) | (6) |
|---|---|---|---|---|---|---|
| | Self-employed | Incorp. entrep. | Self-employed | Incorp. entrep. | Self-employed | Incorp. entrep. |
| Ambiguity aversion (Ellsberg urns) | -0.018 (0.011) | -0.002 (0.003) | | | | |
| Ambig. aversion (belief-hedging) | | | -0.001 (0.008) | -0.011*** (0.004) | 0.007 (0.009) | -0.011*** (0.004) |
| Ambiguity sensitivity | | | | | 0.022*** (0.008) | 0.005 (0.004) |
| N | 523 | 523 | 1147 | 1147 | 1147 | 1147 |
| Log likelihood | -150.114 | -37.097 | -297.873 | -92.690 | -294.097 | -91.861 |
| Mean dep. variable | 0.084 | 0.013 | 0.072 | 0.017 | 0.072 | 0.017 |

*Notes*: Average marginal effects obtained from Probit estimations. Age above 20 and below 65. Columns (1) and (2) use a measure of ambiguity aversion obtained from Ellsberg urns experiments in the LISS in 2010. Columns (3) to (6) use our main measures of ambiguity attitudes using the belief-hedging method based on the stock-market experiments in the LISS in 2018-2021. They sequentially build up our main model: Columns (3) and (4) only include ambiguity aversion, Columns (5) and (6) add ambiguity sensitivity, and by further adding the control variables, we arrive at the full models shown in Columns (1) and (2) of Table 3 in the main paper. Robust standard errors are shown in parentheses. ***/**/* indicate statistical significance at the 1%/5%/10% level. *Source*: Own calculations based on the LISS.

**Table B4: Multinomial Logit Model estimations – coefficients**

| Choice: | Hired manager | Self-employed | Incorp. entrepreneur |
|---|---|---|---|
| Ambiguity aversion | -0.094 | 0.153 | -0.654** |
| | (0.082) | (0.160) | (0.280) |
| Ambiguity sensitivity | 0.079 | 0.376*** | 0.042 |
| | (0.087) | (0.137) | (0.264) |
| Risk aversion | -0.172** | -0.492*** | -1.209*** |
| | (0.082) | (0.140) | (0.254) |
| Optimism | 0.202** | -0.162 | -0.027 |
| | (0.080) | (0.130) | (0.332) |
| Cognitive skill | 0.031 | 0.146 | 0.949** |
| | (0.095) | (0.170) | (0.419) |
| Upper sec. educ. | -0.216 | 0.111 | -0.041 |
| | (0.250) | (0.397) | (1.315) |
| Tertiary education | -0.101 | 0.287 | 1.116 |
| | (0.265) | (0.401) | (1.251) |
| Age | -0.009 | 0.037*** | 0.046* |
| | (0.007) | (0.011) | (0.025) |
| Female | -0.790*** | 0.246 | -1.918* |
| | (0.161) | (0.269) | (1.091) |
| Married | -0.046 | 0.273 | 0.533 |
| | (0.169) | (0.277) | (0.529) |
| No. of children | 0.205*** | 0.166 | -0.076 |
| | (0.071) | (0.123) | (0.235) |
| Constant | -0.271 | -4.758*** | -7.685*** |
| | (0.443) | (0.730) | (1.531) |
| N | 1043 | | |
| Log likelihood | -838.148 | | |

*Notes*: The table shows coefficients obtained from one estimation of a Multinomial Logit Model. We model four choice categories: hired manager, self-employment, incorporated entrepreneur, and non-managerial employee (omitted base category). Age above 20 and below 65. The first six independent variables are standardized. Robust standard errors are shown in parentheses. ***/**/* indicate statistical significance at the 1%/5%/10% level. *Source*: Own calculations based on the LISS.

**Table B5: Further robustness checks – average marginal effects**

| Dep. variable: | (1) Excluding necessity entrepreneurs from the outcome definitions | (2) | (3) Excluding other entrepreneurs from the control groups | (4) | (5) Excluding on-call employees and temp-staffers from the sample | (6) |
|---|---|---|---|---|---|---|
| | Self-employed | Incorp. entrep. | Self-employed | Incorp. entrep. | Self-employed | Incorp. entrep. |
| Ambiguity aversion | 0.009 | -0.010** | 0.012 | -0.011** | 0.011 | -0.010** |
| | (0.009) | (0.004) | (0.009) | (0.005) | (0.010) | (0.004) |
| Ambiguity sensitivity | 0.019** | 0.000 | 0.023*** | 0.002 | 0.020** | 0.000 |
| | (0.008) | (0.004) | (0.009) | (0.004) | (0.009) | (0.004) |
| Risk aversion | -0.026*** | -0.016*** | -0.028*** | -0.020*** | -0.028*** | -0.018*** |
| | (0.008) | (0.005) | (0.009) | (0.006) | (0.009) | (0.005) |
| Optimism | -0.011 | -0.002 | -0.015* | -0.001 | -0.016* | -0.001 |
| | (0.008) | (0.004) | (0.008) | (0.004) | (0.008) | (0.004) |
| Cognitive skill | 0.009 | 0.014** | 0.008 | 0.016** | 0.005 | 0.015** |
| | (0.010) | (0.006) | (0.010) | (0.007) | (0.010) | (0.007) |
| Upper sec. educ. | 0.006 | -0.004 | 0.016 | -0.004 | 0.020 | -0.005 |
| | (0.023) | (0.013) | (0.025) | (0.014) | (0.026) | (0.014) |
| Tertiary education | 0.009 | 0.011 | 0.025 | 0.014 | 0.023 | 0.013 |
| | (0.023) | (0.012) | (0.026) | (0.013) | (0.026) | (0.013) |
| Age | 0.002*** | 0.001** | 0.003*** | 0.001** | 0.002*** | 0.001** |
| | (0.001) | (0.000) | (0.001) | (0.000) | (0.001) | (0.000) |
| Female | 0.030* | -0.014** | 0.030* | -0.015** | 0.029* | -0.016** |
| | (0.016) | (0.006) | (0.017) | (0.006) | (0.017) | (0.006) |
| Married | 0.017 | 0.009 | 0.018 | 0.010 | 0.015 | 0.010 |
| | (0.016) | (0.007) | (0.016) | (0.008) | (0.017) | (0.008) |
| No. of children | 0.007 | -0.004 | 0.007 | -0.003 | 0.008 | -0.003 |
| | (0.007) | (0.003) | (0.007) | (0.004) | (0.007) | (0.004) |
| N | 1062 | 1062 | 1044 | 988 | 1004 | 1004 |
| Log likelihood | -237.789 | -62.708 | -249.653 | -62.436 | -238.702 | -65.132 |
| Mean dep. variable | 0.064 | 0.016 | 0.071 | 0.018 | 0.070 | 0.018 |

*Notes*: Average marginal effects obtained from Probit estimations. Age above 20 and below 65. In Columns (1) and (2), we redefine the outcome variables as self-employed (incorporated entrepreneurs, respectively) who were not unemployed before, indicating that they are not necessity entrepreneurs. In Columns (3) and (4), incorporated entrepreneurs are excluded from the sample in the model estimating the probability of self-employment, and the self-employed are excluded when estimating the probability of being an incorporated entrepreneur. In Columns (5) and (6), on-call employees and temp-staffers are excluded from the sample. The first six independent variables are standardized. Robust standard errors are shown in parentheses. ***/**/* indicate statistical significance at the 1%/5%/10% level. *Source*: Own calculations based on the LISS.